\documentclass[aps,nofootinbib,twocolumn,floatfix,showpacs,preprintnumbers]{revtex4-1}
\usepackage{epsfig,amssymb,color}
\newcommand{\be}{\begin{equation}}
\newcommand{\ee}{\end{equation}}
\newcommand{\bea}{\begin{eqnarray}}
\newcommand{\eea}{\end{eqnarray}}

\def\lsim{\mathrel{\raise.3ex\hbox{$<$\kern-.75em\lower1ex\hbox{$\sim$}}}}
\def\gsim{\mathrel{\raise.3ex\hbox{$>$\kern-.75em\lower1ex\hbox{$\sim$}}}}

\def\d{{\rm d}}

\def\alt{\raise0.3ex\hbox{$\;<$\kern-0.75em\raise-1.1ex\hbox{$\sim\;$}}}
\def\agt{\raise0.3ex\hbox{$\;>$\kern-0.75em\raise-1.1ex\hbox{$\sim\;$}}}
\begin{document}
 \title{No Indications of Axion-Like Particles From Fermi}
\author{Alexander V. Belikov}
\affiliation{Department of Physics, The University of Chicago, Chicago, IL~~60637-1433}
\author{Lisa Goodenough}
\affiliation{Center for Cosmology and Particle Physics, Department of Physics, New York University, New York, NY~~10003}
\author{Dan Hooper}
\affiliation{Center for Particle Astrophysics, Fermi National
Accelerator Laboratory, Batavia, IL~~60510-0500}
\affiliation{Department of Astronomy and Astrophysics, The University of Chicago, Chicago, IL~~60637-1433}

\date{\today}

\begin{abstract}
As very high energy ($\gsim 100$ GeV) gamma rays travel over cosmological distances, their flux is attenuated through interactions with the extragalactic background light. Observations of distant gamma ray sources at energies between $\sim$200 GeV and a few TeV by ground-based gamma-ray telescopes such as HESS, however, have motivated the possibility that the universe is more transparent to very high energy photons than had been anticipated.  One proposed explanation for this is the existence of axion-like-particles (ALPs) which gamma rays can efficiently oscillate into, enabling them to travel cosmological distances without attenuation.  In this article, we use a state-of-the-art model for the extragalactic background light (which is somewhat lower at $\sim \mu$m wavelengths than in previous models) and data from the Fermi Gamma Ray Space Telescope to calculate the spectra at 1-100 GeV of two gamma-ray sources, 1ES1101-232 at redshift $z=0.186$ and H2356-309 at $z=0.165$, in conjunction with the measurements of ground-based telescopes, to test the ALP hypothesis. We find that these observations can be well-fit by an intrinsic power-law source spectrum with indices of -1.72 and -2.1 for 1ES1101-232 and H2356-309, respectively, and that no ALPs or other exotic physics is necessary to explain the observed degree of attenuation. While this does not exclude the possibility that ALPs are involved in the cosmological propagation of gamma rays, it does reduce the motivation for such new physics.

\end{abstract}

\preprint{FERMILAB-PUB-10-247-A}
\maketitle

\section{introduction}\label{sec:intro}

Very high energy gamma rays ($E_{\gamma} \gsim 100$ GeV) scattering with photons in the infrared to ultraviolet range can exceed the threshold for electron-positron pair production.  As a consequence, gamma rays are predicted to experience significant attenuation over cosmological distances, leading to the suppression of the gamma-ray flux from high redshift sources such as active galactic nuclei and gamma-ray bursts~\cite{Stecker:1992wi}.

Beginning several years ago, however, ground-based atmospheric Cerenkov telescopes (HESS, MAGIC, VERITAS) began to report measurements of distant gamma-ray sources which did not appear to exhibit as much attenuation as had been generally anticipated~\cite{Aharonian:2005gh,Mazin:2007pn}. In particular, the HESS collaboration reported spectra from two relatively high redshift sources, the active galactic nuclei (AGN) H2356-309 at $z=0.186$ and 1ES1101-232 at $z=0.165$, that appeared to exhibit relatively little attenuation. These observations indicate that either these particular sources are injecting a very hard spectrum of gamma rays ($dN_{\gamma}/dE_{\gamma} \propto E_{\gamma}^{-\Gamma}$ with $\Gamma \sim -0.6$ to 0.2, in contrast to $\Gamma \sim 2$ as predicted from second order Fermi acceleration), or that the universe is considerably more transparent to very high energy photons than had been expected~\cite{Stecker:2007jq, Stecker:2008fp}. If the latter is the case, it could indicate a low density of extragalactic background light (EBL) or, alternatively, the existence of exotic physics which enables very high energy photons to traverse over cosmological distances without experiencing attenuation. 

One possibility that has received some attention is that very high energy photons may be oscillating into axion-like particles (ALPs), which do not experience significant attenuation, only to be reconverted into photons before reaching Earth. In Ref.~\cite{Hooper:2007bq} (and later Ref.~\cite{Hochmuth:2007hk}) it was proposed that the magnetic fields in and around the very high energy gamma-ray sources could facilitate the efficient conversion of very high energy gamma rays into ALPs. This effect would be expected to become significant at the GeV-TeV energy range currently being studied by satellite and ground-based gamma-ray telescopes. Furthermore, it is plausible that a significant fraction of such particles could be reconverted into photons through interactions with the magnetic field of the Milky Way~\cite{Simet:2007sa}. Alternatively, it was suggested in Ref.~\cite{DeAngelis:2007dy} that photon-ALP conversions could take place through interactions with the (nano-gauss scale) intergalactic magnetic field.  Another explanation consistent with both the EBL calculations and AGN models involves ultra high energy gamma rays of energies below 50 EeV crossing cosmological distances and interacting with the EBL close to Earth, thus generating secondary photons that are observed by gamma-ray telescopes~\cite{Essey:2009zg, Essey:2009ju}. 

The measurements of the spectra of cosmologically distant gamma-rays sources by ground-based atmospheric Cerenkov telescopes have thus far been unable to distinguish between the various possible explanations for the lack of observed attenuation in these sources. Studies of these sources' spectra over a wider range of energies, however, could be used to constrain their initial (unattenuated) spectral shapes, and to determine with greater precision the degree to which very high energy gamma rays are attenuated by the EBL. In particular, because the EBL has a negligible density at wavelengths below 0.1 $\mu$m, there is no significant attenuation of gamma rays with energies below the corresponding threshold energy of $\sim$$20$ GeV, and only modest attenuation below $\sim$$100$ GeV.

The most significant source of uncertainty in determining the attenuation of gamma rays is the EBL spectrum. The two complementary methods of estimating the EBL spectrum are the integrated galaxy counts, dependent on star formation history, and the absolute measurements, prone to uncertainties due to foreground contamination. As it stands the model proposed by Franceschini et al. \cite{Franceschini:2008tp} provides the minimal spectrum for the first peak in the optical to near infrared (NIR) and corresponds to minimal values of the calculated total EBL ~\cite{Horiuchi:2008jz}.
In this article, we use publicly available data from the Large Area Telescope (LAT) on the Fermi Gamma Ray Space Telescope to study the spectrum of two active galactic nuclei (1ES1101-232 and H2356-309) between 1 GeV and 100 GeV. When combined with the higher energy spectral measurements of ground-based gamma-ray telescopes, a more complete picture is obtained, covering more than three decades of energy, including a range for which there is expected to be little or no attenuation due to electron-positron pair production. We find that these measurements imply a relatively low density of the EBL, but that the observed spectra can be accounted for without the presence of ALPs or other exotic physics.

\section{The Attenuation of Cosmological Gamma-Ray Sources}
\label{sec:propagation}

As very high energy gamma rays propagate through the universe, their interactions with the EBL and the resulting production of electron-positron pairs cause their spectrum to become attenuated by a factor of $e^{-\tau(E_{\gamma},z_0)}$, where the optical depth, $\tau(E_{\gamma},z_0)$, is given by

\begin{equation}
\tau(E_{\gamma},z_0)= \int_0^{z_0} \frac{\d z}{(1+z)H(z)}\int \d\omega \frac{\d n_{\rm EBL}}{\d\omega}(\omega,z)\,\bar\sigma(E_{\gamma},\omega,z)\,
\label{eqn:attenuation}
\end{equation}
with
\begin{equation}
\bar\sigma  (E_{\gamma},\omega,z) = \!\int_{-1}^{1-2(m_e c^2)^2\!/(\omega E_{\gamma})}\frac{1}{2} (1-\mu)\sigma_{\gamma\gamma}(E_{\gamma},\omega,\mu)\,\d\mu \,.
\label{eqn:cross-section}
\end{equation}
Here, $z_0$ is the redshift of the source, $H(z)$ is the rate of Hubble expansion, $E_{\gamma}$ and $\omega$ are the (appropriately redshifted)
source and background photon energies, respectively, $\mu$ is the
cosine of the angle between the incoming and target photon, $\sigma_{\gamma\gamma}$ is the cross section for electron-positron pair production, and $\d n_{\rm EBL}/ \d\omega$ denotes the spectrum and density of the EBL~\cite{Gould:1967zzb}.

For a gamma ray of energy $E_{\gamma}$, the pair production cross section is largest for background photons of energy

\begin{equation}
\omega_{\rm max} \simeq 2(m_e c^2)^2/E_{\gamma} \simeq 0.5 \biggl (\frac{1 \text{TeV}}{E_{\gamma}} \biggr ) \: \mbox{eV},
\label{eqn:maxphotonenergy}
\end{equation}
or equivalently 

\begin{equation}
\lambda_{\rm max} \simeq 1.24 \,\mu\mbox{m} \times \bigg(\frac{E_{\gamma}}{1\,{\rm TeV}}\bigg).
\label{eqn:maxphotonwavelength}
\end{equation}
From this expression, we see that gamma rays with $E_\gamma = 100$ GeV (approximately the highest energies measured by Fermi) interact most efficiently with the far-UV component of the EBL.  Lower energy gamma rays, constituting the bulk of the Fermi events, experience relatively little attenuation through pair production, since the EBL is negligible at far-UV and shorter wavelengths.  In contrast, the HESS measurements of gamma rays at $\sim$ 2 TeV directly probe the EBL in the near-IR and longer wavelengths.

\begin{figure}[t]
\begin{center}
\includegraphics[width=.48\textwidth]{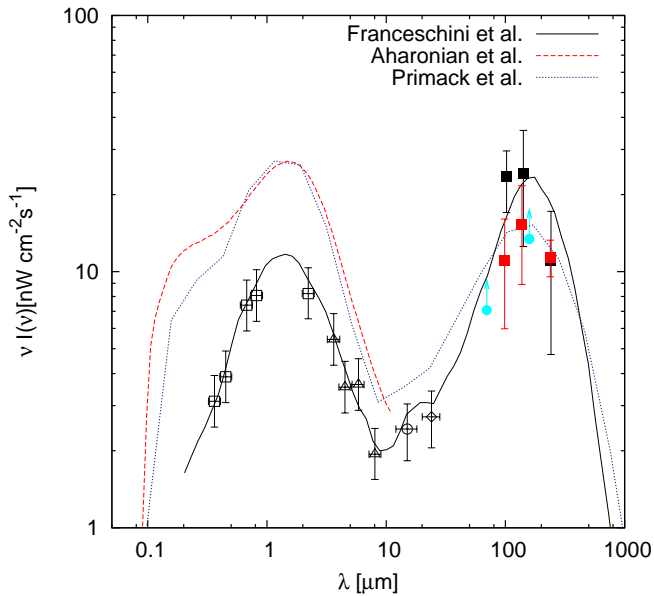}
\end{center}
\caption{Representative models of the extragalactic background light (EBL) compared to various measurements.  The solid black line is the model of Franceschini {\it et al.}~\cite{Franceschini:2008tp}, which we adopt in our calculations. The dashed red line denotes the model used by the HESS collaboration in Ref.~\cite{Aharonian:2005gh}. The dotted blue line is the model of Primack {\it et al.}, using the Kennicutt {\it et al.} stellar initial mass function~\cite{Primack:2000xp,Kennicutt:1983mv}. The data are as described in Ref.~\cite{Franceschini:2008tp}. The three rightmost data points in the far-IR are from the analysis of DIRBE data by Madau {\it et al}. (black squares)~\cite{Madau:1999yh} and the re-analysis of that data by Lagache {\it et al}. (red squares)~\cite{Lagache:1999ji}. The two lower limits given at 70 and 160 $\mu$m (cyan filled circles) come from a stacking analysis of Spitzer data described in Ref.~\cite{Dole:2006de}. The mid-IR data points at 15 and 24 $\mu$m are the resolved fraction of the EBL by the deep ISO surveys IGTES~\cite{Elbaz:2002vd,Papovich:2004vh}. The near-IR data point at 8 $\mu$m (open triangle) is from Ref.~\cite{Franceschini:2008tp}, while the three lower wavelength points (open triangles) are from Ref.~\cite{Fazio:2004mu}, using the integration of the IRAC galaxy counts. The far-UV/optical/near-IR points from 0.25 $\mu$m to 2.5 $\mu$m (open squares) are from an analysis of Ref.~\cite{Madau:1999yh} based on deep galaxy counts.
}
\label{fig:IREBL}
\end{figure}

The EBL consists of the photons emitted by extragalactic sources, both resolved and unresolved, over all cosmic epochs. Its density and spectrum are difficult to measure directly due to the presence of bright atmospheric and galactic foregrounds, including Zodiacal scattered light, interplanetary dust emission, Galactic starlight, and high Galactic latitude cirrus emission (Galactic stellar light reflected by high latitude dust).  In Fig.~\ref{fig:IREBL}, we show three examples of EBL models, as described in Refs.~\cite{Franceschini:2008tp,Aharonian:2005gh,Primack:2000xp}.

The distribution of extragalactic optical and infrared background photons used in this paper is that from Franceschini {\it et al}.~\cite{Franceschini:2008tp}.  Their analysis of the EBL uses survey data in the optical, near-IR, and sub-millimeter from ground-based observatories as well as multi-wavelength information from Hubble Space Telescope (HST), Infrared Space Observatory (ISO), and Spitzer Space Telescope, while also utilizing direct measurements of and upper limits on the EBL by COBE to place additional constraints on their estimate.  The Franceschini {\it et al}. model of the optical background is based on HST deep galaxy count estimates. To avoid the large near-IR excess found in some analyses based on the COBE measurements at 1.2 and 3.5 $\mu$m, and the accompanying strong discontinuity in the EBL at $\sim 1 \mu$m, they use a model of the near-IR background derived from direct integration of the ultra-deep HST and Spitzer counts.

\begin{figure}
\begin{center}
\includegraphics[width=.48\textwidth]{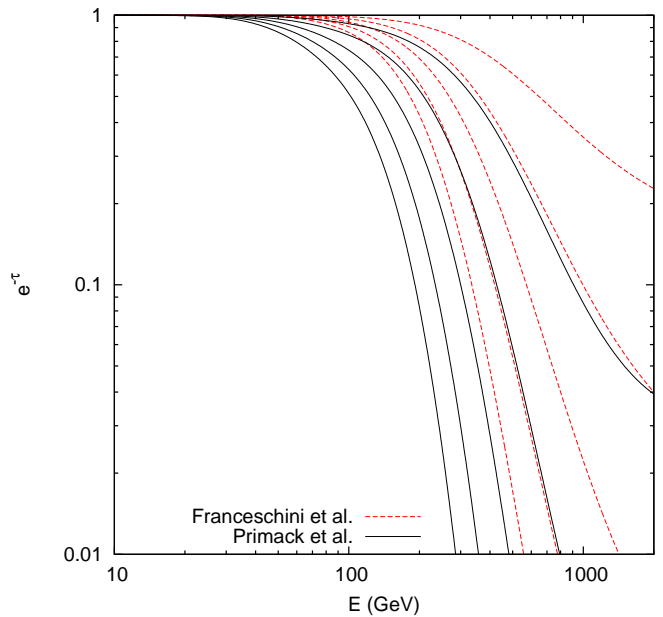}
\end{center}
\caption{The attenuation factor for gamma rays due to electron-positron pair production for sources at redshifts of $z=0.1$, $z=0.2$, $z=0.3$, $z=0.4$, and $z=0.5$, from right-to-left. The dashed and solid curves denote the attenuation for the EBL models of Franceschini {\it et al.}~\cite{Franceschini:2008tp} and Primack {\it et al.}~\cite{Primack:2000xp,Kennicutt:1983mv}, respectively (see Fig.~\ref{fig:IREBL}).}
\label{fig:attenuation}
\end{figure}

Referring to Fig.~\ref{fig:IREBL}, we point out that the model of Franceschini {\it et al.}~\cite{Franceschini:2008tp} has a considerably lower estimate for the EBL's density in the near-IR, and a somewhat higher estimate in the far-IR, relative to the other models shown. Franceschini {\it et al.}~\cite{Franceschini:2008tp} constrain the near-IR at 8 $\mu$m using Spitzer IRAC observations of the faint number counts~\cite{Fazio:2004kx,Lonsdale:2003pg} and the local galaxy luminosity function~\cite{Huang:2007dq}. These data significantly constrain the EBL at the junction between the stellar photospheric emission in the optical/near-IR and the far-IR dust emission peak.

The most relevant yet
controversial region of EBL for the attenuation of gamma rays of energies of 100-500 GeV is the range from of optical through near IR.  The analysis by Bernstein et al.~\cite{Bernstein:2001rz} obtained background values a factor of $\sim 3$ greater than those estimated from direct HST counts. Their approach was based on subtraction the contributions of the Zodiacal light and the diffuse galactic light from the total signal measured by HST. Such a discrepancy would imply that either there exists a truly diffuse background of unknown origin or that the HST missed a large part of the light from the faint galaxies. 
It was later  demonstrated that the original analysis strongly underestimated the systematic errors and that the background estimates have to be considered as upper limits~\cite{Bernstein:2007bb}.

Throughout our analysis, we assume that the shape and normalization of the background radiation does not change except for redshifting between the source and the observer.  As suggested by the models of Ref.~\cite{Kneiske:2002wi}, this is a reasonable assumption for $z<0.4$.  In Fig.~\ref{fig:attenuation}, we plot the attenuation resulting from this EBL model, as a function of observed energy, and for sources at various redshifts.  The energy dependence of the optical depth leads to modifications in the source spectrum as measured locally.  In general, even a weak dependence on energy of the optical depth can lead to large changes in the source spectrum because of the exponential dependence of the attenuation.  Since the optical depth is generally larger for photons with higher energies, the effect of attenuation is to make the locally observed spectrum softer than the source spectrum, as well as to reduce the overall intensity.

\section{The Modification of Gamma-Ray Spectra by Axion-Like Particles}
\label{sec:ALPs}
As discussed in Refs.~\cite{Hooper:2007bq,Hochmuth:2007hk}, the probability of a photon converting into an ALP in a magnetic field is related to the size of the domain of the magnetic field and the strength of the magnetic field component along the polarization vector of the photon.  Given the typical sizes of astrophysical accelerators and the magnetic field strengths present in them, significant photon-to-ALP conversion can occur in or near very high energy gamma-ray sources over a large range of allowed ALP parameter space.  For an unpolarized photon source, one expects $1/3$ of the original photons to be converted into ALPs at the source above a critical energy, ${\cal E}$.  That is, for energies $E_{\gamma}\gg{\cal E}$ there could potentially be an ALP flux from very high energy gamma-ray sources as large as $\sim$50\% of the residual gamma-ray flux exiting the source.  The critical energy, ${\cal E}$, can naturally fall in the gamma-ray band.  The photon-ALP mixture then propagates toward the Milky Way without significant further oscillations, since in the absence of an external electromagnetic field, the ALPs and the photons effectively decouple.  During propagation over cosmological distances, the spectrum of photons is depleted via pair production.  As a result, upon reaching the Milky Way, the very high energy flux will be dominated by ALPs.

The reconversion probability of ALPs into photons in the the Galactic Magnetic Field depends on a number of factors, including the ALP-photon coupling, the ALP mass, the photon energy, and the strength and geometry of the magnetic field.  The CERN Axion Solar Telescope (CAST) experiment~\cite{Andriamonje:2007ew,Zioutas:2004hi} provides a direct bound on the ALP-photon coupling of $g_{a \gamma} \lesssim 8.8 \times 10^{-11} \text{GeV}^{-1}$ for $m_a\alt 0.02\,$eV.  For ultra-light ALPs ($m_a\alt 10^{-11}\,$eV), the absence of gamma rays from SN~1987A yields a limit of $g_{a \gamma} \lesssim 1 \times 10^{-11} \text{GeV}^{-1}$ \cite{Brockway:1996yr} or even $g_{a \gamma} \lesssim 0.3 \times 10^{-11} \text{GeV}^{-1}$ \cite{Grifols:1996id}.  However, in the range of $10^{-11}\,{\rm eV}\ll m_a\ll 10^{-2}\,$eV, the CAST bound is the most general and stringent.  The optimal range of parameters for the reconversion mechanism is $10^{-10}\alt m_a\alt 10^{-8}$, over which couplings as large as $g_{a \gamma} \approx 8 \times 10^{-11} \text{GeV}^{-1}$ are consistent with present bounds. Although the characteristics of the Galactic Magnetic Field are not very well known, especially in the directions toward the Galactic Center or away from the Galactic Plane, reasonable models of the Galactic Magnetic Field predict regions of the sky in which the reconversion probability can be 20\% or larger for an ALP-photon coupling of $g_{a} = 5 \times 10^{-11} \text{GeV}^{-1}$~\cite{Simet:2007sa}. Moreover, Ref.~\cite{Simet:2007sa} found that, in these models, a significant fraction of the sky corresponds to a probability larger than 10\%.  Thus, appreciable reconversion probabilities are possible, though difficult to predict given the scarcity of our knowledge regarding the structure of the Galactic Magnetic Field. 

We note that for the most viable range of parameters for which the ALP mechanism is efficient ($g_{a} \approx 5 \times 10^{-11} \text{GeV}^{-1}$, $m_a\sim10^{-9}\,$eV and microgauss-scale fields, for example), the critical energy, ${\cal E}$, above which the probability of a photon converting into an ALP is of the order unity, falls in the sub-GeV range.  Therefore, the observed gamma-ray spectrum will not contain peculiar features resulting from the onset of photon-ALP oscillations.  Thus, over the whole range of energies observed by Fermi and atmospheric Cherenkov telescopes, the ALP flux generated at the source via oscillation will have the same spectrum as the photons at the source.

\begin{figure}
\begin{center}
\includegraphics[width=.47\textwidth]{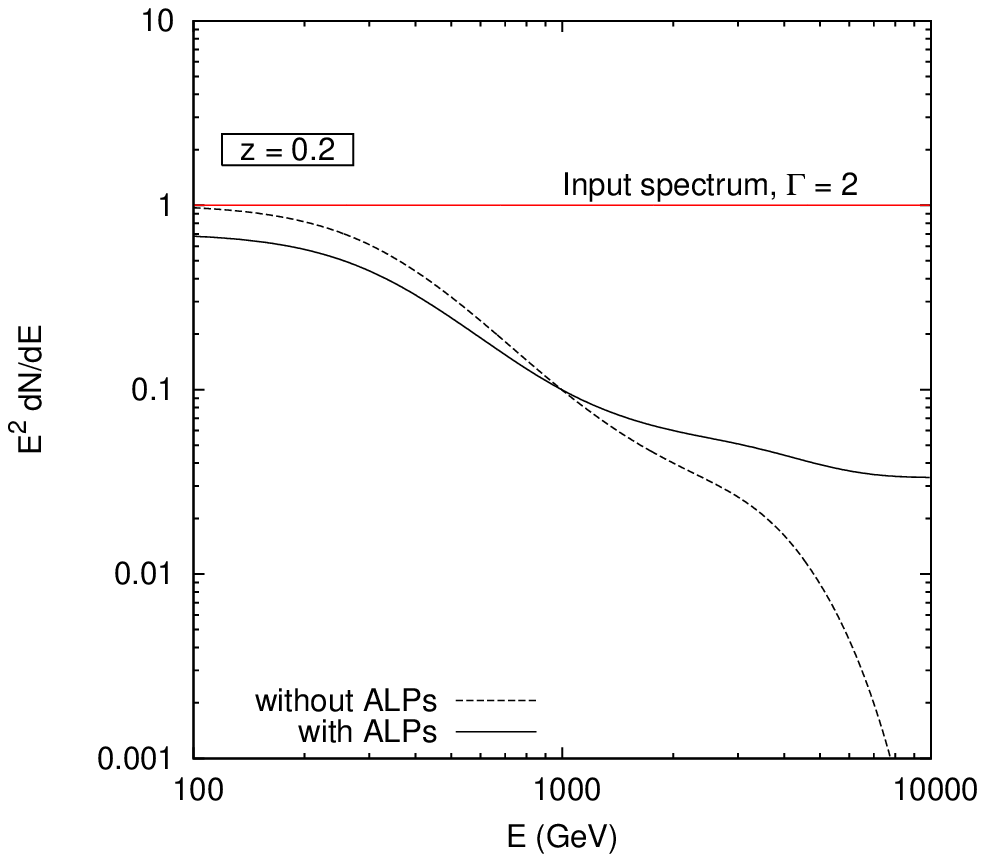}\\
\includegraphics[width=.47\textwidth]{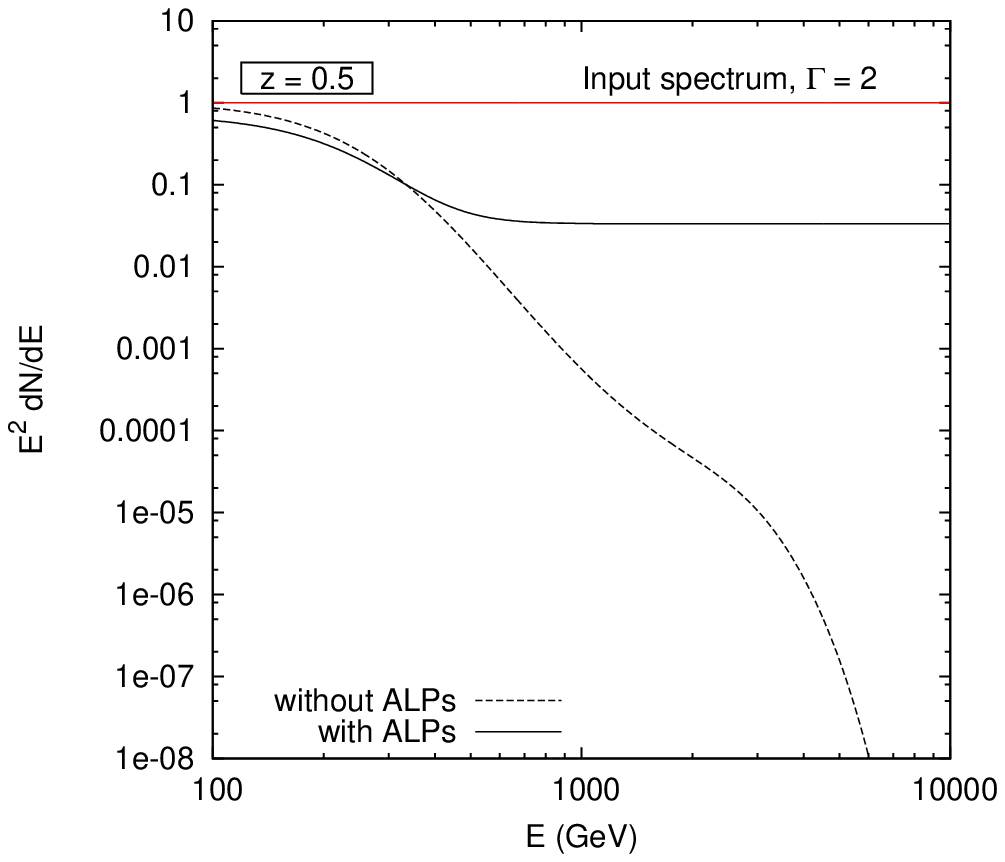}\\
\end{center}
\caption{The gamma-ray spectrum at Earth from a source with an injected spectrum of $dN/dE \propto E^{-2}$, after propagation over a distance of $z=0.2$ (top) and $z=0.5$ (bottom). Results are shown with and without the effects of photon-ALP oscillations. The ALP-photon mixing mitigates the impact of absorption via pair production and leads to a plateau in the spectrum at high energies. We have calculated the effects of ALP-photon mixing assuming a source conversion probability of 1/3 and a Milky Way reconversion probability of 0.1.  The vertical axis is in arbitrary units.}
\label{fig:comparisonspectra}
\end{figure}

In Fig.~\ref{fig:comparisonspectra}, we show the effects of ALPs on the gamma-ray spectrum at Earth for the case of a source spectrum of $dN/dE \propto E^{-\Gamma}$ with $\Gamma=2$ after propagating from $z=0.2$ (top) and $z=0.5$ (bottom), assuming oscillation into ALPs at the source with a probability of 1/3 and reconversion in the Galactic Magnetic Field with a probability of 0.1.  For comparison, we also include the results for a standard scenario with no ALPs. Clearly, in the presence of ALP-photon mixing, the quasi-exponential cutoff is interrupted, leading the spectrum to plateau at high energies.

\section{The Measured Source Spectra}

The optical depth of the universe to very high energy photons increases with redshift, so the spectra of very distant sources experience greater attenuation.  Any mechanism mitigating the attenuation is therefore most apparent in the spectra of high redshift sources. In our analysis, we consider two relatively high redshift gamma-ray sources which have each been observed by both HESS~\cite{Aharonian:2005gh} and Fermi~\cite{Abdo:2009ct}, 1ES1101-232 and H2356-309. These sources are both blazars, and were first detected at TeV energies by HESS in 2006. 1ES1101-232 is located at a distance of $z=0.186$ at $(l,b)=(273.19^\circ,33.08^\circ)$.  H2356-309 is located at a distance of $z=0.165$ at $(l,b)=(12.86^\circ,-78.04^\circ)$. HESS has reported data on the spectra of these sources in the energy range $\sim200 \: \text{GeV} < E < 3 \:\text{TeV}$ \cite{Aharonian:2005gh}. Fermi has not detected variability in either of these sources~\cite{Abdo:2009ct}.

The Fermi-LAT team made their first year gamma-ray event data publicly available at the end of August 2009, and this data is now updated daily on their website~\cite{fermidata}.  Additionally, Fermi has supplied the Fermi Science Tools for the analysis of LAT data \cite{fermitools}.  In our analysis we use the LAT event data collected over the period from August 4, 2008 to January 11, 2010.  

The LAT point spread function (PSF) for front-converting events\footnote{The front, or ``thin'', section of the LAT tracker has twelve tungsten converters, each of which is 0.035\% radiation lengths thick, while the back, or ``thick'', section has four 0.18\% radiation length tungsten converters.} is a decreasing function of photon energy with a 95\% containment radius of $\sim$$1.2^\circ$ at 1 GeV and less than $0.25^\circ$ above 10 GeV.  Back-converting events have a PSF of $\sim$$2^\circ$ at 1 GeV and less than $0.5^\circ$ above 10 GeV.  We note that an analysis by Ref.~\cite{Neronov:2010bi} suggests that the width of the PSF for back-converting photons with energies above $\sim$$10$ GeV may be underestimated in the Monte Carlo simulations used to generate the LAT response functions. To prevent contamination of the spectra by other nearby sources, we limit the angular size of our analysis region according to a radius equal to half of the angular distance to the nearest known source. For 1ES1101-232 and H2356-309, the closest known sources are $239.151^\prime$ and $125.244^\prime$ away, respectively.  The energy dependence of the PSF thus determines the minimum energy in the analysis of the spectrum.  For consistency we calculate the spectra for both AGN over the same energy range, 1-100 GeV.

\begin{figure}[!htb]
\begin{tabular}{c}
\epsfig{file=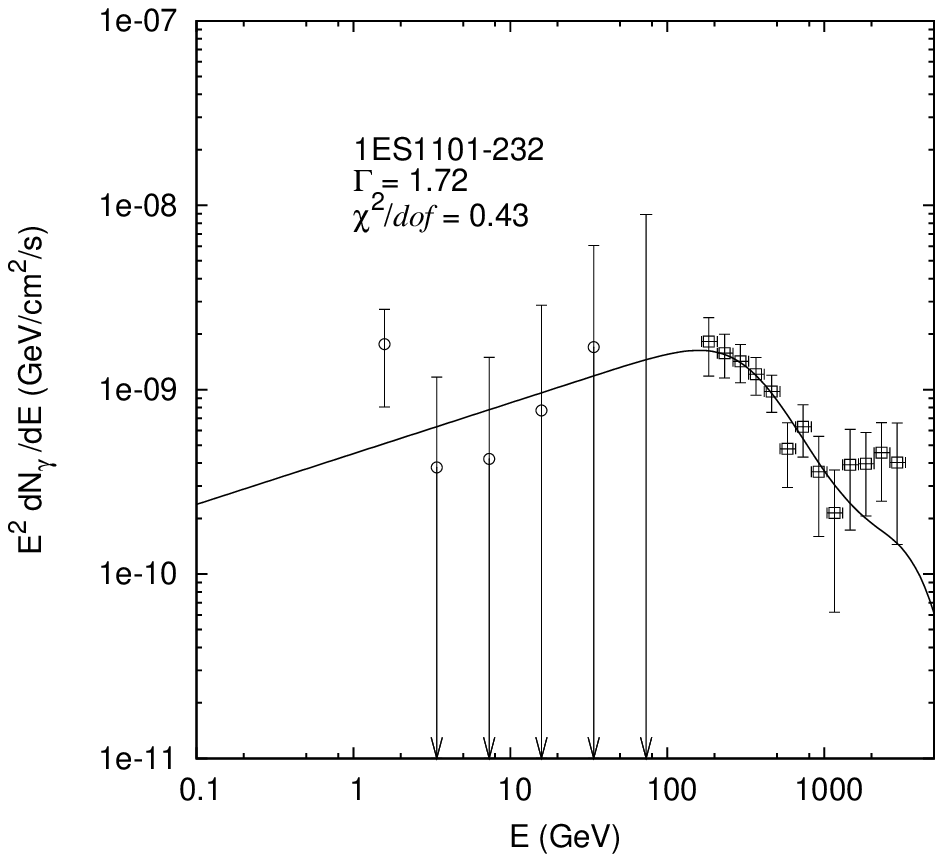,width=1.0\columnwidth, angle=0} \\
\\
\epsfig{file=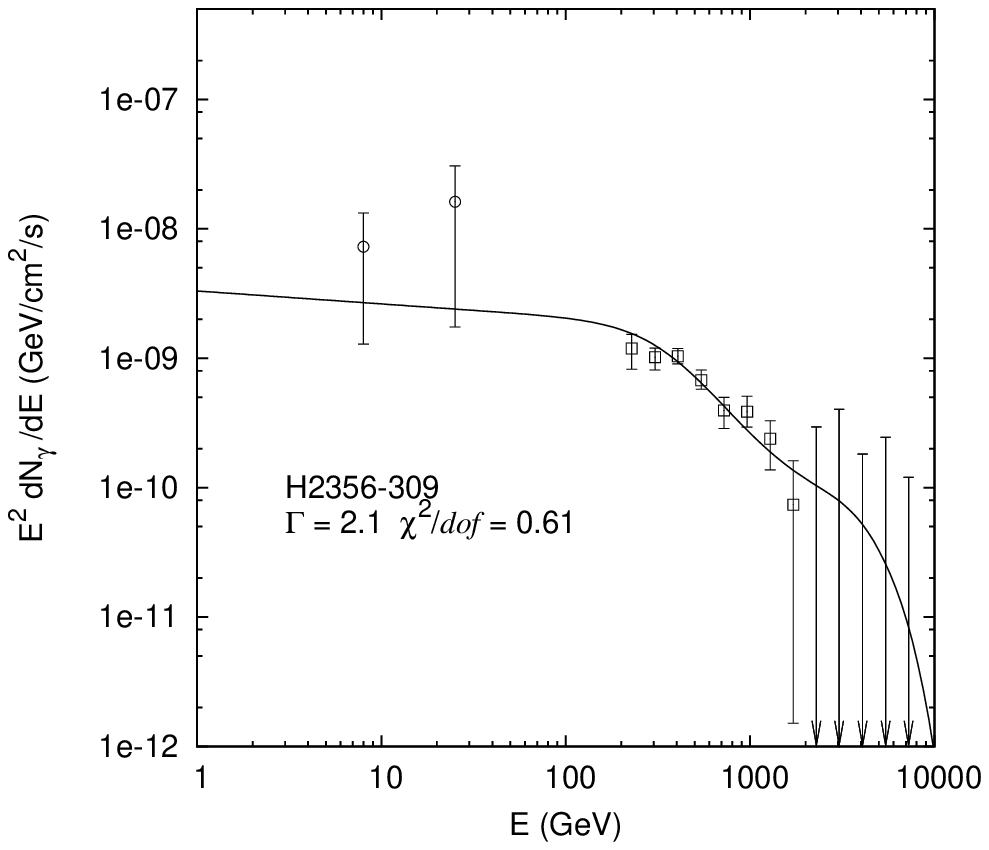,width=1.0\columnwidth, angle=0}
\end{tabular}
\caption{The spectra of gamma-ray sources 1ES1101-232 at $z=0.186$ (upper) and H2356-309 at $z=0.165$ (lower) as determined from Fermi data (\textit{black circles}) and as measured by HESS (\textit{black squares}).  The curves shown are the calculated best fits to the data assuming absorption of gamma rays via electron-positron pair production interactions with the EBL.  The intrinsic source spectra are taken to be of the form $\d N/\d E \propto E^{-\Gamma}$.  The best-fit values for the power law indices are $\Gamma = 1.72$ for 1ES1101-232 and $\Gamma =  2.1$ for H2356-309.  The EBL model used is that of Franceschini {\it et al}.~\cite{Franceschini:2008tp}. See text for more details.}
\label{fig:sourcespectra}
\end{figure}

To determine the spectra of the AGN, we calculate the total flux and subtract off the background.  We estimate the background spectrum using two methods, which we refer to as the ``Fermi'' method and the ``annulus'' method.  For the Fermi method, we assume that the background flux is the sum of the diffuse galactic emission~\cite{diffusemodel} and the isotropic emission~\cite{Abdo:2010nz} (which is also referred to as the extragalactic diffuse emission) as measured by the LAT team.  The diffuse flux has its source in cosmic rays interacting with the interstellar gas and interstellar radiation field, while the isotropic flux is composed mainly of gamma rays from unresolved extragalactic sources.  For the annulus method, we calculate the total flux in an annulus of width 0.25$^{\circ}$, centered around the source, and with an inner radius that is 10\% larger than the PSF at a given energy. 

These two methods for calculating the background flux in the region of 1ES1101-232 give results that agree very well. However, for H2356-309, the Fermi and the annulus backgrounds do not agree; the annulus background is more than an order of magnitude greater than the Fermi background at $\sim$2 GeV, and exceeds the Fermi background by even larger factors at higher energies.  We suspect that this can be explained by the existence of a currently unresolved point source in the region and/or by contamination from the nearby source 1FGL J2350.1-3005 due to the underestimation of the PSF suggested by Ref.~\cite{Neronov:2010bi}, as mentioned earlier in this section. The Fermi spectra shown in the upper frame of Fig.~\ref{fig:sourcespectra} (1ES1101-232) was calculated using the Fermi background method, while the results in the lower frame (H2356-309) have made use of the annulus method. The error bars shown for the spectra in Fig.~\ref{fig:sourcespectra} were calculated assuming statistical uncertainties in the total flux and systematic+statistical uncertainties (though systematics dominate) in the isotropic flux.

\section{Results}
\label{results}

In Fig.~\ref{fig:sourcespectra}, we compare the spectra of 1ES1101-232 and H2356-309 as observed by Fermi and HESS to the shape predicted after propagation, assuming an intrinsic source spectrum of the form $\d N_{\gamma}/\d E \propto E^{-\Gamma}$ and using the EBL model of Ref.~\cite{Franceschini:2008tp} (see Fig.~\ref{fig:IREBL}). We find that in the case of  1ES1101-232, the combination of Fermi and HESS data is best-fit for a value of $\Gamma=1.72$ ($\chi^2/dof \approx 0.43$). For H2356-309, a value of $\Gamma=2.1$ yields an acceptable fit to the spectrum above $\sim$3 GeV ($\chi^2/dof \approx 0.61$). One should keep in mind that the injected spectrum from astrophysical sources need not follow a simple power-law~\cite{Stecker:2009eq,Inoue:1996vv}, although power-law forms do yield good fits to the observed spectra in the cases considered here. On this basis, we conclude that the observed spectra of these sources do not contain compelling indications of axion-like particles.

A few comments are in order at this time. In Ref.~\cite{Simet:2007sa}, a similar calculation, for the same two gamma-ray sources analyzed here, 1ES1101-232 and H2356-309, was performed using HESS (but not Fermi) data and the EBL model of Ref.~\cite{Primack:2000xp}. In their analysis, they found that, under standard astrophysical assumptions ({\it ie.}~no ALPS), the intrinsic source spectrum of 1ES1101-232 was required to have a very hard power-law index of $\Gamma \approx -0.2$.  Similarly, they found that H2356-309 required a power law index of $\Gamma \approx 0.6$ in the absence of ALPs.  The large discrepancy between the values of $\Gamma$ calculated by Ref.~\cite{Simet:2007sa} (-0.2 and 0.6) and the values found in this study (1.72 and 2.1) is largely due to the difference in the EBL models used.  For gamma rays with energies in the HESS range, the dominant contributors to pair production are EBL photons with wavelengths of $3.75 \: \mu\text{m} > \lambda > 0.25 \: \mu\text{m}$.  The density of the EBL in the model used in our calculations~\cite{Franceschini:2008tp} is a factor of $\sim$$2-3$ times smaller than that used in Ref.~\cite{Simet:2007sa} (see Fig.~\ref{fig:IREBL}).  Thus, the predicted attenuation is much less and the required intrinsic source spectrum much softer than that found in Ref.~\cite{Simet:2007sa}.  Moreover, the Fermi data at energies of $\sim$1-100 GeV confirms this conclusion by constraining the spectrum in a energy range largely unaffected by attenuation.

A similar analysis was done in Ref.~\cite{Stecker:2007zj}, where 1ES1101-232 and H2356-309 were analyzed, among other gamma-ray sources, using HESS but not Fermi data.  They used two EBL models of Ref.~\cite{Stecker:2005qs} corresponding to different galaxy evolution models, a baseline model (B) and a fast evolution model (FE).  The calculated intrinsic power-law index of 1ES1101-232 in their analysis was found to be $\Gamma \approx 1.0$ using model B and $\Gamma \approx 1.5$ using model FE.  For H2356-309 the intrinsic power-law index was found to be $\Gamma \approx 1.5$ and $\Gamma \approx 1.9$, for models B and FE, correspondingly.  These values are much closer than those of Ref.~\cite{Simet:2007sa} to our present results, which can explained by the similarities of the EBL models.

\bigskip
\bigskip

\section{Summary and Conclusions}

In this article, we have studied the spectra of two gamma-ray sources, 1ES1101-232 at redshift $z=0.186$ and H2356-309 at $z=0.165$, based on the observations of the the Fermi Gamma Ray Space Telescope and the ground-based telescope HESS. While the observations of HESS at energies above a few hundred GeV indicate a somewhat surprising degree of the transparency of the universe to very high energy photons, the reason for this had not until now been clear. By including in our analysis recent data at lower energies ($\lsim$100 GeV) from Fermi, we find that the spectra of 1ES1101-232 and H2356-309 are consistent with an intrinsic source spectrum with power law indices of approximately -1.7 and -2.1, respectively. The lack of attenuation observed by HESS can be accounted for if the density of the extragalactic background light (EBL) is somewhat low in the UV-optical-near-IR range (such as that in the model of Franceschini {\it et al.}). 

Having emphasized the importance of the EBL model we conclude that no exotic physics, such as oscillations between photons and axion-like particles (ALPs), is required to explain the relative lack of attenuation observed by HESS.

\section*{Acknowledgments}
We would like to thank Joel Primack, Martin Raue and Shunsaku Horiuchi for insightful comments and valuable discussions.  This work has been supported by the US Department of Energy, including grant DE-FG02-95ER40896, and by NASA grant NAG5-10842. LG is supported by DOE OJI grant \# DE-FG02-06ER41417.

\bibliography{AGN}

\begin{thebibliography}{42}
\expandafter\ifx\csname natexlab\endcsname\relax\def\natexlab#1{#1}\fi
\expandafter\ifx\csname bibnamefont\endcsname\relax
  \def\bibnamefont#1{#1}\fi
\expandafter\ifx\csname bibfnamefont\endcsname\relax
  \def\bibfnamefont#1{#1}\fi
\expandafter\ifx\csname citenamefont\endcsname\relax
  \def\citenamefont#1{#1}\fi
\expandafter\ifx\csname url\endcsname\relax
  \def\url#1{\texttt{#1}}\fi
\expandafter\ifx\csname urlprefix\endcsname\relax\def\urlprefix{URL }\fi
\providecommand{\bibinfo}[2]{#2}
\providecommand{\eprint}[2][]{\url{#2}}

\bibitem[{\citenamefont{Stecker et~al.}(1992)\citenamefont{Stecker, de~Jager,
  and Salamon}}]{Stecker:1992wi}
\bibinfo{author}{\bibfnamefont{F.~W.} \bibnamefont{Stecker}},
  \bibinfo{author}{\bibfnamefont{O.~C.} \bibnamefont{de~Jager}},
  \bibnamefont{and} \bibinfo{author}{\bibfnamefont{M.~H.}
  \bibnamefont{Salamon}}, \bibinfo{journal}{Astrophys. J.}
  \textbf{\bibinfo{volume}{390}}, \bibinfo{pages}{L49} (\bibinfo{year}{1992}).

\bibitem[{\citenamefont{Aharonian et~al.}(2006)}]{Aharonian:2005gh}
\bibinfo{author}{\bibfnamefont{F.}~\bibnamefont{Aharonian}}
  \bibnamefont{et~al.} (\bibinfo{collaboration}{H.E.S.S.}),
  \bibinfo{journal}{Nature} \textbf{\bibinfo{volume}{440}},
  \bibinfo{pages}{1018} (\bibinfo{year}{2006}), \eprint{astro-ph/0508073}.

\bibitem[{\citenamefont{Mazin and Raue}(2007)}]{Mazin:2007pn}
\bibinfo{author}{\bibfnamefont{D.}~\bibnamefont{Mazin}} \bibnamefont{and}
  \bibinfo{author}{\bibfnamefont{M.}~\bibnamefont{Raue}},
  \bibinfo{journal}{Astron. Astrophys.} \textbf{\bibinfo{volume}{471}},
  \bibinfo{pages}{439} (\bibinfo{year}{2007}), \eprint{astro-ph/0701694}.

\bibitem[{\citenamefont{Stecker and Scully}(2007)}]{Stecker:2007jq}
\bibinfo{author}{\bibfnamefont{F.~W.} \bibnamefont{Stecker}} \bibnamefont{and}
  \bibinfo{author}{\bibfnamefont{S.~T.} \bibnamefont{Scully}}
  (\bibinfo{year}{2007}), \eprint{0710.2252}.

\bibitem[{\citenamefont{Stecker and
  Scully}(2009{\natexlab{a}})}]{Stecker:2008fp}
\bibinfo{author}{\bibfnamefont{F.~W.} \bibnamefont{Stecker}} \bibnamefont{and}
  \bibinfo{author}{\bibfnamefont{S.~T.} \bibnamefont{Scully}},
  \bibinfo{journal}{Astrophys. J. Lett.} \textbf{\bibinfo{volume}{691}},
  \bibinfo{pages}{L91} (\bibinfo{year}{2009}{\natexlab{a}}),
  \eprint{0807.4880}.

\bibitem[{\citenamefont{Hooper and Serpico}(2007)}]{Hooper:2007bq}
\bibinfo{author}{\bibfnamefont{D.}~\bibnamefont{Hooper}} \bibnamefont{and}
  \bibinfo{author}{\bibfnamefont{P.~D.} \bibnamefont{Serpico}},
  \bibinfo{journal}{Phys. Rev. Lett.} \textbf{\bibinfo{volume}{99}},
  \bibinfo{pages}{231102} (\bibinfo{year}{2007}), \eprint{0706.3203}.

\bibitem[{\citenamefont{Hochmuth and Sigl}(2007)}]{Hochmuth:2007hk}
\bibinfo{author}{\bibfnamefont{K.~A.} \bibnamefont{Hochmuth}} \bibnamefont{and}
  \bibinfo{author}{\bibfnamefont{G.}~\bibnamefont{Sigl}},
  \bibinfo{journal}{Phys. Rev.} \textbf{\bibinfo{volume}{D76}},
  \bibinfo{pages}{123011} (\bibinfo{year}{2007}), \eprint{0708.1144}.

\bibitem[{\citenamefont{Simet et~al.}(2008)\citenamefont{Simet, Hooper, and
  Serpico}}]{Simet:2007sa}
\bibinfo{author}{\bibfnamefont{M.}~\bibnamefont{Simet}},
  \bibinfo{author}{\bibfnamefont{D.}~\bibnamefont{Hooper}}, \bibnamefont{and}
  \bibinfo{author}{\bibfnamefont{P.~D.} \bibnamefont{Serpico}},
  \bibinfo{journal}{Phys. Rev.} \textbf{\bibinfo{volume}{D77}},
  \bibinfo{pages}{063001} (\bibinfo{year}{2008}), \eprint{0712.2825}.

\bibitem[{\citenamefont{De~Angelis et~al.}(2007)\citenamefont{De~Angelis,
  Mansutti, and Roncadelli}}]{DeAngelis:2007dy}
\bibinfo{author}{\bibfnamefont{A.}~\bibnamefont{De~Angelis}},
  \bibinfo{author}{\bibfnamefont{O.}~\bibnamefont{Mansutti}}, \bibnamefont{and}
  \bibinfo{author}{\bibfnamefont{M.}~\bibnamefont{Roncadelli}},
  \bibinfo{journal}{Phys. Rev.} \textbf{\bibinfo{volume}{D76}},
  \bibinfo{pages}{121301} (\bibinfo{year}{2007}), \eprint{0707.4312}.

\bibitem[{\citenamefont{Essey and Kusenko}(2010)}]{Essey:2009zg}
\bibinfo{author}{\bibfnamefont{W.}~\bibnamefont{Essey}} \bibnamefont{and}
  \bibinfo{author}{\bibfnamefont{A.}~\bibnamefont{Kusenko}},
  \bibinfo{journal}{Astropart. Phys.} \textbf{\bibinfo{volume}{33}},
  \bibinfo{pages}{81} (\bibinfo{year}{2010}), \eprint{0905.1162}.

\bibitem[{\citenamefont{Essey et~al.}(2010)\citenamefont{Essey, Kalashev,
  Kusenko, and Beacom}}]{Essey:2009ju}
\bibinfo{author}{\bibfnamefont{W.}~\bibnamefont{Essey}},
  \bibinfo{author}{\bibfnamefont{O.~E.} \bibnamefont{Kalashev}},
  \bibinfo{author}{\bibfnamefont{A.}~\bibnamefont{Kusenko}}, \bibnamefont{and}
  \bibinfo{author}{\bibfnamefont{J.~F.} \bibnamefont{Beacom}},
  \bibinfo{journal}{Phys. Rev. Lett.} \textbf{\bibinfo{volume}{104}},
  \bibinfo{pages}{141102} (\bibinfo{year}{2010}), \eprint{0912.3976}.

\bibitem[{\citenamefont{Franceschini et~al.}(2008)\citenamefont{Franceschini,
  Rodighiero, and Vaccari}}]{Franceschini:2008tp}
\bibinfo{author}{\bibfnamefont{A.}~\bibnamefont{Franceschini}},
  \bibinfo{author}{\bibfnamefont{G.}~\bibnamefont{Rodighiero}},
  \bibnamefont{and} \bibinfo{author}{\bibfnamefont{M.}~\bibnamefont{Vaccari}},
  \bibinfo{journal}{Astronomy and Astrophysics} \textbf{\bibinfo{volume}{487}},
  \bibinfo{pages}{837} (\bibinfo{year}{2008}), \eprint{0805.1841}.

\bibitem[{\citenamefont{Horiuchi et~al.}(2009)\citenamefont{Horiuchi, Beacom,
  and Dwek}}]{Horiuchi:2008jz}
\bibinfo{author}{\bibfnamefont{S.}~\bibnamefont{Horiuchi}},
  \bibinfo{author}{\bibfnamefont{J.~F.} \bibnamefont{Beacom}},
  \bibnamefont{and} \bibinfo{author}{\bibfnamefont{E.}~\bibnamefont{Dwek}},
  \bibinfo{journal}{Phys. Rev.} \textbf{\bibinfo{volume}{D79}},
  \bibinfo{pages}{083013} (\bibinfo{year}{2009}), \eprint{0812.3157}.

\bibitem[{\citenamefont{Gould and Schreder}(1967)}]{Gould:1967zzb}
\bibinfo{author}{\bibfnamefont{R.~J.} \bibnamefont{Gould}} \bibnamefont{and}
  \bibinfo{author}{\bibfnamefont{G.~P.} \bibnamefont{Schreder}},
  \bibinfo{journal}{Phys. Rev.} \textbf{\bibinfo{volume}{155}},
  \bibinfo{pages}{1404} (\bibinfo{year}{1967}).

\bibitem[{\citenamefont{Primack et~al.}(2001)\citenamefont{Primack, Somerville,
  Bullock, and Devriendt}}]{Primack:2000xp}
\bibinfo{author}{\bibfnamefont{J.~R.} \bibnamefont{Primack}},
  \bibinfo{author}{\bibfnamefont{R.~S.} \bibnamefont{Somerville}},
  \bibinfo{author}{\bibfnamefont{J.~S.} \bibnamefont{Bullock}},
  \bibnamefont{and} \bibinfo{author}{\bibfnamefont{J.~E.~G.}
  \bibnamefont{Devriendt}}, \bibinfo{journal}{AIP Conf. Proc.}
  \textbf{\bibinfo{volume}{558}}, \bibinfo{pages}{463} (\bibinfo{year}{2001}),
  \eprint{astro-ph/0011475}.

\bibitem[{\citenamefont{Kennicutt}(1983)}]{Kennicutt:1983mv}
\bibinfo{author}{\bibfnamefont{R.~C.} \bibnamefont{Kennicutt},
  \bibfnamefont{Jr.}}, \bibinfo{journal}{Astrophys. J.}
  \textbf{\bibinfo{volume}{272}}, \bibinfo{pages}{54} (\bibinfo{year}{1983}).

\bibitem[{\citenamefont{Madau and Pozzetti}(2000)}]{Madau:1999yh}
\bibinfo{author}{\bibfnamefont{P.}~\bibnamefont{Madau}} \bibnamefont{and}
  \bibinfo{author}{\bibfnamefont{L.}~\bibnamefont{Pozzetti}},
  \bibinfo{journal}{MNRAS} \textbf{\bibinfo{volume}{312}}, \bibinfo{pages}{L9}
  (\bibinfo{year}{2000}), \eprint{astro-ph/9907315}.

\bibitem[{\citenamefont{Lagache et~al.}(2000)\citenamefont{Lagache, Haffner,
  Reynolds, and Tufte}}]{Lagache:1999ji}
\bibinfo{author}{\bibfnamefont{G.}~\bibnamefont{Lagache}},
  \bibinfo{author}{\bibfnamefont{L.~M.} \bibnamefont{Haffner}},
  \bibinfo{author}{\bibfnamefont{R.~J.} \bibnamefont{Reynolds}},
  \bibnamefont{and} \bibinfo{author}{\bibfnamefont{S.~L.} \bibnamefont{Tufte}},
  \bibinfo{journal}{Astron. Astrophys.} \textbf{\bibinfo{volume}{354}},
  \bibinfo{pages}{247} (\bibinfo{year}{2000}), \eprint{astro-ph/9911355}.

\bibitem[{\citenamefont{Dole et~al.}(2006)}]{Dole:2006de}
\bibinfo{author}{\bibfnamefont{H.}~\bibnamefont{Dole}} \bibnamefont{et~al.},
  \bibinfo{journal}{Astron. Astrophys.} \textbf{\bibinfo{volume}{451}},
  \bibinfo{pages}{417} (\bibinfo{year}{2006}), \eprint{astro-ph/0603208}.

\bibitem[{\citenamefont{Elbaz et~al.}(2002)}]{Elbaz:2002vd}
\bibinfo{author}{\bibfnamefont{D.}~\bibnamefont{Elbaz}} \bibnamefont{et~al.},
  \bibinfo{journal}{Astron. Astrophys.} \textbf{\bibinfo{volume}{384}},
  \bibinfo{pages}{848} (\bibinfo{year}{2002}), \eprint{astro-ph/0201328}.

\bibitem[{\citenamefont{Papovich et~al.}(2004)}]{Papovich:2004vh}
\bibinfo{author}{\bibfnamefont{C.}~\bibnamefont{Papovich}}
  \bibnamefont{et~al.}, \bibinfo{journal}{Astrophys. J. Suppl.}
  \textbf{\bibinfo{volume}{154}}, \bibinfo{pages}{70} (\bibinfo{year}{2004}),
  \eprint{astro-ph/0406035}.

\bibitem[{\citenamefont{Fazio et~al.}(2004{\natexlab{a}})}]{Fazio:2004mu}
\bibinfo{author}{\bibfnamefont{G.~G.} \bibnamefont{Fazio}} \bibnamefont{et~al.}
  (\bibinfo{collaboration}{The IRAC}), \bibinfo{journal}{Astrophys. J. Suppl.}
  \textbf{\bibinfo{volume}{154}}, \bibinfo{pages}{10}
  (\bibinfo{year}{2004}{\natexlab{a}}), \eprint{astro-ph/0405616}.

\bibitem[{\citenamefont{Fazio et~al.}(2004{\natexlab{b}})}]{Fazio:2004kx}
\bibinfo{author}{\bibfnamefont{G.~G.} \bibnamefont{Fazio}}
  \bibnamefont{et~al.}, \bibinfo{journal}{Astrophys. J. Suppl.}
  \textbf{\bibinfo{volume}{154}}, \bibinfo{pages}{39}
  (\bibinfo{year}{2004}{\natexlab{b}}), \eprint{astro-ph/0405595}.

\bibitem[{\citenamefont{Lonsdale et~al.}(2003)}]{Lonsdale:2003pg}
\bibinfo{author}{\bibfnamefont{C.~J.} \bibnamefont{Lonsdale}}
  \bibnamefont{et~al.}, \bibinfo{journal}{Publ. Astron. Soc. Pac.}
  \textbf{\bibinfo{volume}{115}}, \bibinfo{pages}{897} (\bibinfo{year}{2003}),
  \eprint{astro-ph/0305375}.

\bibitem[{\citenamefont{Huang et~al.}(2007)}]{Huang:2007dq}
\bibinfo{author}{\bibfnamefont{J.~S.} \bibnamefont{Huang}}
  \bibnamefont{et~al.}, \bibinfo{journal}{Astrophys. J.}
  \textbf{\bibinfo{volume}{664}}, \bibinfo{pages}{840} (\bibinfo{year}{2007}),
  \eprint{0704.3609}.

\bibitem[{\citenamefont{Bernstein et~al.}(2002)\citenamefont{Bernstein,
  Freedman, and Madore}}]{Bernstein:2001rz}
\bibinfo{author}{\bibfnamefont{R.~A.} \bibnamefont{Bernstein}},
  \bibinfo{author}{\bibfnamefont{W.~L.} \bibnamefont{Freedman}},
  \bibnamefont{and} \bibinfo{author}{\bibfnamefont{B.~F.}
  \bibnamefont{Madore}}, \bibinfo{journal}{Astrophys.J.}
  \textbf{\bibinfo{volume}{571}}, \bibinfo{pages}{56} (\bibinfo{year}{2002}),
  \eprint{astro-ph/0112153}.

\bibitem[{\citenamefont{Bernstein}(2007)}]{Bernstein:2007bb}
\bibinfo{author}{\bibfnamefont{R.~A.} \bibnamefont{Bernstein}},
  \bibinfo{journal}{Astrophys.J.} \textbf{\bibinfo{volume}{666}},
  \bibinfo{pages}{663} (\bibinfo{year}{2007}).

\bibitem[{\citenamefont{Kneiske et~al.}(2002)\citenamefont{Kneiske, Mannheim,
  and Hartmann}}]{Kneiske:2002wi}
\bibinfo{author}{\bibfnamefont{T.~M.} \bibnamefont{Kneiske}},
  \bibinfo{author}{\bibfnamefont{K.}~\bibnamefont{Mannheim}}, \bibnamefont{and}
  \bibinfo{author}{\bibfnamefont{D.~H.} \bibnamefont{Hartmann}}
  (\bibinfo{year}{2002}), \eprint{astro-ph/0202104}.

\bibitem[{\citenamefont{Andriamonje et~al.}(2007)}]{Andriamonje:2007ew}
\bibinfo{author}{\bibfnamefont{S.}~\bibnamefont{Andriamonje}}
  \bibnamefont{et~al.} (\bibinfo{collaboration}{CAST}), \bibinfo{journal}{JCAP}
  \textbf{\bibinfo{volume}{0704}}, \bibinfo{pages}{010} (\bibinfo{year}{2007}),
  \eprint{hep-ex/0702006}.

\bibitem[{\citenamefont{Zioutas et~al.}(2005)}]{Zioutas:2004hi}
\bibinfo{author}{\bibfnamefont{K.}~\bibnamefont{Zioutas}} \bibnamefont{et~al.}
  (\bibinfo{collaboration}{CAST}), \bibinfo{journal}{Phys. Rev. Lett.}
  \textbf{\bibinfo{volume}{94}}, \bibinfo{pages}{121301}
  (\bibinfo{year}{2005}), \eprint{hep-ex/0411033}.

\bibitem[{\citenamefont{Brockway et~al.}(1996)\citenamefont{Brockway, Carlson,
  and Raffelt}}]{Brockway:1996yr}
\bibinfo{author}{\bibfnamefont{J.~W.} \bibnamefont{Brockway}},
  \bibinfo{author}{\bibfnamefont{E.~D.} \bibnamefont{Carlson}},
  \bibnamefont{and} \bibinfo{author}{\bibfnamefont{G.~G.}
  \bibnamefont{Raffelt}}, \bibinfo{journal}{Phys. Lett.}
  \textbf{\bibinfo{volume}{B383}}, \bibinfo{pages}{439} (\bibinfo{year}{1996}),
  \eprint{astro-ph/9605197}.

\bibitem[{\citenamefont{Grifols et~al.}(1996)\citenamefont{Grifols, Masso, and
  Toldra}}]{Grifols:1996id}
\bibinfo{author}{\bibfnamefont{J.~A.} \bibnamefont{Grifols}},
  \bibinfo{author}{\bibfnamefont{E.}~\bibnamefont{Masso}}, \bibnamefont{and}
  \bibinfo{author}{\bibfnamefont{R.}~\bibnamefont{Toldra}},
  \bibinfo{journal}{Phys. Rev. Lett.} \textbf{\bibinfo{volume}{77}},
  \bibinfo{pages}{2372} (\bibinfo{year}{1996}), \eprint{astro-ph/9606028}.

\bibitem[{\citenamefont{Abdo et~al.}(2009)}]{Abdo:2009ct}
\bibinfo{author}{\bibfnamefont{A.~A.} \bibnamefont{Abdo}} \bibnamefont{et~al.}
  (\bibinfo{collaboration}{Fermi LAT}), \bibinfo{journal}{Astrophys. J.}
  \textbf{\bibinfo{volume}{707}}, \bibinfo{pages}{1310} (\bibinfo{year}{2009}),
  \eprint{0910.4881}.

\bibitem[{fer({\natexlab{a}})}]{fermidata}
\bibinfo{howpublished}{\url{http://fermi.gsfc.nasa.gov/ssc/data/}}.

\bibitem[{fer({\natexlab{b}})}]{fermitools}
\bibinfo{howpublished}{\url{http://fermi.gsfc.nasa.gov/ssc/data/analysis/}}.

\bibitem[{\citenamefont{Neronov et~al.}(2010)\citenamefont{Neronov, Semikoz,
  Tinyakov, and Tkachev}}]{Neronov:2010bi}
\bibinfo{author}{\bibfnamefont{A.}~\bibnamefont{Neronov}},
  \bibinfo{author}{\bibfnamefont{D.~V.} \bibnamefont{Semikoz}},
  \bibinfo{author}{\bibfnamefont{P.~G.} \bibnamefont{Tinyakov}},
  \bibnamefont{and} \bibinfo{author}{\bibfnamefont{I.~I.}
  \bibnamefont{Tkachev}} (\bibinfo{year}{2010}), \eprint{1006.0164}.

\bibitem[{dif()}]{diffusemodel}
\bibinfo{howpublished}{\url{http://fermi.gsfc.nasa.gov/ssc/data/access/lat/Bac%
kgroundModels.html}}.

\bibitem[{\citenamefont{Abdo et~al.}(2010)}]{Abdo:2010nz}
\bibinfo{author}{\bibfnamefont{A.~A.} \bibnamefont{Abdo}} \bibnamefont{et~al.}
  (\bibinfo{collaboration}{The Fermi-LAT}), \bibinfo{journal}{Phys. Rev. Lett.}
  \textbf{\bibinfo{volume}{104}}, \bibinfo{pages}{101101}
  (\bibinfo{year}{2010}), \eprint{1002.3603}.

\bibitem[{\citenamefont{Stecker and
  Scully}(2009{\natexlab{b}})}]{Stecker:2009eq}
\bibinfo{author}{\bibfnamefont{F.~W.} \bibnamefont{Stecker}} \bibnamefont{and}
  \bibinfo{author}{\bibfnamefont{S.~T.} \bibnamefont{Scully}}
  (\bibinfo{year}{2009}{\natexlab{b}}), \eprint{0911.3659}.

\bibitem[{\citenamefont{Inoue and Takahara}(1996)}]{Inoue:1996vv}
\bibinfo{author}{\bibfnamefont{S.}~\bibnamefont{Inoue}} \bibnamefont{and}
  \bibinfo{author}{\bibfnamefont{F.}~\bibnamefont{Takahara}},
  \bibinfo{journal}{Astrophys.J.} \textbf{\bibinfo{volume}{463}},
  \bibinfo{pages}{555} (\bibinfo{year}{1996}).

\bibitem[{\citenamefont{Stecker et~al.}(2007)\citenamefont{Stecker, Baring, and
  Summerlin}}]{Stecker:2007zj}
\bibinfo{author}{\bibfnamefont{F.~W.} \bibnamefont{Stecker}},
  \bibinfo{author}{\bibfnamefont{M.~G.} \bibnamefont{Baring}},
  \bibnamefont{and} \bibinfo{author}{\bibfnamefont{E.~J.}
  \bibnamefont{Summerlin}}, \bibinfo{journal}{Astrophys. J.}
  \textbf{\bibinfo{volume}{667}}, \bibinfo{pages}{L29} (\bibinfo{year}{2007}),
  \eprint{0707.4676}.

\bibitem[{\citenamefont{Stecker et~al.}(2006)\citenamefont{Stecker, Malkan, and
  Scully}}]{Stecker:2005qs}
\bibinfo{author}{\bibfnamefont{F.~W.} \bibnamefont{Stecker}},
  \bibinfo{author}{\bibfnamefont{M.~A.} \bibnamefont{Malkan}},
  \bibnamefont{and} \bibinfo{author}{\bibfnamefont{S.~T.}
  \bibnamefont{Scully}}, \bibinfo{journal}{Astrophys. J.}
  \textbf{\bibinfo{volume}{648}}, \bibinfo{pages}{774} (\bibinfo{year}{2006}),
  \eprint{astro-ph/0510449}.

\end{thebibliography}
\bibliographystyle{apsrev}

\end{document}